\documentclass[aps,prb,groupedaddress,amsmath,amssymb,twocolumn]{revtex4}
	\usepackage{dcolumn}
	\usepackage{bm}
	\usepackage{graphicx}
	\usepackage{dcolumn}
\newcommand\beq{\begin{equation}}
\newcommand\eeq{\end{equation}}
\newcommand\beqa{\begin{eqnarray}}
\newcommand\eeqa{\end{eqnarray}}
\newcommand{\nn}{\nonumber\\}
\begin{document}


\title{Equation of state of a seven-dimensional hard-sphere fluid.
{Percus--Yevick theory and molecular dynamics simulations}}


\author{Miguel Robles}
\email[]{mrp@cie.unam.mx}
\author{Mariano L\'opez de Haro}
\email[]{malopez@servidor.unam.mx}
\affiliation{Centro de Investigaci\'on en
Energ\'{\i}a, UNAM, Temixco, Morelos 62580, M{e}xico}
\author{Andr\'es Santos}
\email[]{andres@unex.es}
\homepage[]{http://www.unex.es/fisteor/andres/}
\affiliation{Departamento de F\'{\i}sica, Universidad de Extremadura,
E-06071 Badajoz, Spain}


\date{\today}

\begin{abstract}
Following the work of Leutheusser [Physica A \textbf{127}, 667 (1984)], the
solution to the Percus--Yevick equation for a seven-dimensional hard-sphere
fluid is explicitly found. This allows the derivation of the equation of
state for the fluid taking both the virial and the compressibility routes.
An analysis of the virial coefficients and the determination of the radius
of convergence of the virial series are carried out. Molecular dynamics
simulations of the same system are also performed and a comparison between
the simulation results for the compressibility factor and theoretical
expressions for the same quantity is presented.

\end{abstract}



\maketitle

\section{Introduction\label{sec1}}
In liquid theory there has been a long lasting
interest on the equilibrium properties of high-dimensional hard-sphere fluids,
 especially in the last few years.\cite{FI81}$^{\text{--}}$\cite{CM03c}
{ Such an interest has arisen from many different sources. To begin with, given
 the relative simplicity of the intermolecular interactions in these hard-core
 systems, they are amenable to both theoretical and computer simulation
 studies. In this sense and as it occurs in other problems in theoretical
 and mathematical physics, it is an asset that one can deal with hard spheres
 in arbitrary dimensionality and exploit some of the features that these
 systems have in common, for instance the fact that they all exhibit a
 first-order freezing transition. Furthermore, and as { conjectured} by
 Frisch and Percus,\cite{FP99} in the case of fluids high spatial
 dimensionality { may have} a parallel with limiting high density situations, { so that by increasing the dimensionality one may obtain
 at least a rough idea of any thermodynamic phenomenology that extends to such
 dimensionality}.  An example of this expectation is the recent investigation of the
 demixing problem in mixtures of hard hyperspheres.\cite{YSH00}}

Computer simulation studies of hard-sphere  fluids in dimensions greater than
three are very scarce. To the best of our knowledge, only the four- and
five-dimensional simple\cite{MT84,LM90} and multicomponent\cite{GAH01}
fluids have been simulated. This is not surprising since the computational
effort needed to obtain reliable results increases significantly with the
dimensionality.

Exact information on the equation of state (EOS) usually comes from the virial
coefficients $B_n$ defined by\cite{HM86}
\beqa
Z\equiv\frac{p}{\rho k_BT}&=&1+\sum_{n=2}^\infty B_n \rho^{n-1}\nn
&=&1+\sum_{n=2}^\infty b_n \eta^{n-1}.
\label{a1}
\eeqa
In this equation,  $p$ is the pressure, $\rho$ is the number density, $k_B$ is
the Boltzmann constant, $T$ is the temperature, and $Z$ is the compressibility
factor. The second virial coefficient is $B_{2}=2^{d-1}v_d\sigma^d$, where
$d$ is the dimensionality, $\sigma$ is the diameter of a sphere, and
$v_d=(\pi /4)^{d/2}/\Gamma (1+d/2)$ is the volume of a $d$-dimensional sphere
of unit diameter. In the second line of Eq.\ (\ref{a1}) we have introduced the
packing fraction
$\eta\equiv \rho v_d\sigma^d$ and the reduced virial coefficients
$b_n\equiv B_n/(v_d\sigma^d)^{n-1}=2^{(d-1)(n-1)}B_n/B_2^{n-1}$.
The radius of convergence of the virial series (\ref{a1}),
$\rho_{\text{conv}}=\lim_{n\to\infty}|B_n/B_{n+1}|$, is the modulus of the
singularity of $Z(\rho)$ closest to the origin in the complex $\rho$ plane. If
such a singularity were located on the positive real axis, then all the virial
coefficients $B_n$ would be positive for large $n$.

The exact expression for the third virial coefficient is\cite{LB82,BC87}
\beq
\frac{B_3}{B_2^2}=2\frac{B_{3/4}(d/2+1,1/2)}{B(d/2+1/2,1/2)},
\label{a2}
\eeq
where $B(a,b)=\Gamma(a)\Gamma(b)/\Gamma(a+b)$ is the beta function and
$B_x(a,b)$ is the incomplete beta function. Both $B_2$ and $B_3$ are positive
definite for arbitrary $d$.
The analytic evaluation of the fourth virial coefficient is much more involved.
Luban and Baram\cite{LB82,J82} derived exact expressions for two of the three
terms contributing to $B_4$ and proposed a semi-empirical formula for the
remaining contribution. More recently, Clisby and McCoy\cite{CM03a,CM03c}
showed that $B_4$ can be evaluated exactly for any \textit{even} dimension
$d$ and gave the explicit results for $d=4$, 6, 8, 10, and 12. They also
computed numerically\cite{CM03b,CM03c} the fifth and sixth virial
coefficients through $d=50$. The results show that, while $B_5$ remains
positive, $B_4$  and $B_6$ become negative for $d\geq 8$ and $d\geq 6$,
respectively.
This suggests the possibility that, even in the three-dimensional case, there
might be negative virial coefficients $B_n$ for sufficiently large $n$.\cite{newnote1}
The fact that the virial coefficients are not positive definite and that they
may have alternate signs is of importance in connection with the radius of
convergence of the virial expansion (\ref{a1}), as mentioned before.

\subsection{{ The scenario for high dimensionality}}
{ The high-dimensionality limit of hard hyperspheres has been the subject of several studies.\cite{FRW85,KF86,FP99,PS00}.   By means of asymptotic methods and
heuristic arguments, Frisch and Percus\cite{FP99} were led to the following scenario in that limit:}
\begin{itemize}
\item
The fourth virial coefficient is negative. Beyond that term,  the virial
expansion is an alternating series.
\item
The virial series is convergent for $\widehat{\rho}<1$, where
$\widehat{\rho}\equiv
2\eta^{1/d}$ is the scaled density per dimension. In terms of the packing
fraction, the virial series converges for $\eta <\eta_{\text{conv}}=2^{-d}$.
\item
In the density range  $\widehat{\rho}<1$ the second virial term dominates over
the remaining ones, so that
\beq
Z\approx 1+B_2\rho =1+\frac{\widehat{\rho}^d}{2}.
\label{9}
\eeq
\item
Even though the virial expansion does not converge for
$\widehat{\rho}>1$ (oscillatory divergence), the truncated series (\ref{9})
remains valid within the interval
$1<\widehat{\rho}<(1-\epsilon)\widehat{\rho}_0$,
where $\epsilon=\mathcal{O}(d^{-1})$ and
\beq
\widehat{\rho}_0=\left(1.148
d^{-1/6}e^{-1.473d^{1/3}}\right)^{-1/d}\sqrt{e/2}\sim \sqrt{e/2}\simeq 1.17.
\label{10}
\eeq
\item
At the density $\widehat{\rho}=\widehat{\rho_0}$ an infinite compressibility
spinodal appears, thus indicating a first-order transition to the
high-dimensional solid.
\end{itemize}
In an independent paper,  Parisi and Slanina\cite{PS00}
reached similar conclusions from a toy model based on simplified HNC equations.
They obtained  that, while in the limit $d\to\infty$ the EOS for
$\widehat{\rho}<1$ is given by Eq.~(\ref{9}), in the interval
$1<\widehat{\rho}<\widehat{\rho}_0=\sqrt{e/2}$, one has
\beq
Z(\widehat{\rho})=1+\widehat{\rho}^d\left[\frac{1}{2}+
\Delta(\widehat{\rho})\right],
\label{14}
\eeq
where
\beq
\Delta(\widehat{\rho})=\left[\frac{e\kappa(\widehat{\rho})}
{2\widehat{\rho}^2}\right]^d
,
\label{15}
\eeq
$\kappa(\widehat{\rho})$ being the solution to\cite{note1}
\beq
\ln\frac{2\widehat{\rho}^2}{e}=\ln\left(1+\sqrt{1-\kappa^2}\right)-\sqrt{1-\kappa^2}.
\label{16}
\eeq
Note that, since $\ln \kappa< \ln\left(1+\sqrt{1-\kappa^2}\right)-\sqrt{1-\kappa^2}$
for $0<\kappa<1$, one has  $\lim_{d\to \infty}\Delta(\widehat{\rho})= 0$
for $\widehat{\rho}<\widehat{\rho}_0$.
Although, strictly speaking, Eq.~(\ref{16}) cannot be extended to
$\widehat{\rho}>\widehat{\rho}_0$,  Eq.~(\ref{15}) suggests that
$\lim_{d\to\infty}\Delta(\widehat{\rho})= \infty$ in that density domain, in
agreement with the phase transition noted by Frisch  and Percus.\cite{note2}

\subsection{{ Approximate equations of state}}
As in two and three dimensions, one can make use of approximate schemes to
represent the EOS of hard hyperspheres.
Several proposals have been made in the literature for the EOS based on the
knowledge of the first few virial coefficients.\cite{BC87,ASV89,SMS89,SM90,
MSAV91} For illustration, we review here a few of them making use of the first
three virial coefficients. The extension to higher virial coefficients is
straightforward.
\subsubsection{{ Truncated virial series}}
The first obvious choice is the truncated virial expansion
\beq
Z_{[2,0]}(\eta)=1+b_2\eta+b_3\eta^2.
\label{x2}
\eeq
As discussed above, this simple approximation becomes more and more accurate
in the stable fluid domain as the dimensionality increases.
In a way analogous to Eq.\ (\ref{x2}) it is possible to define a truncated expansion $Z_{[n,0]}$ from the knowledge of the first $n+1$ virial coefficients.
\subsubsection{{ Pad\'e approximants}}
One can also construct Pad\'e approximants of the form $Z_{[n,m]}$ from the first $n+m+1$ virial coefficients. For instance,
\beq
Z_{[1,1]}(\eta)=\frac{b_2+(b_2^2-b_3)\eta}{b_2-b_3\eta},
\label{x3}
\eeq
\beq
Z_{[0,2]}(\eta)=\left[1-b_2\eta+(b_2^2-b_3)\eta^2\right]^{-1}.
\label{x4}
\eeq
\subsubsection{{ Colot--Baus approximation}}
Colot and Baus\cite{CB86,BC87} proposed  (truncated) \textit{rescaled} virial
expansions, where the  series expansion of $(1-\eta)^dZ(\eta)$, rather than that of  $Z(\eta)$, is truncated. Let us denote by  $Z_{[n,0]}^{\text{BC}}(\eta)$ the truncated rescaled virial expansion that makes use of the first $n+1$ virial coefficients. For example,
\beq
Z_{[2,0]}^{\text{BC}}(\eta)=\frac{1+(b_2-d)\eta+[b_3-b_2 d+d(d-1)/2]\eta^2}
{(1-\eta)^d}.
\label{x5}
\eeq
The pole of order $d$ at the (unphysical) packing fraction $\eta=1$ is suggested by the
scaled particle theory.
\subsubsection{{ Maeso--Solana--Amor\'os--Villar approximation}}
Maeso et al.\cite{MSAV91} combined the advantages of Pad\'e approximants and
rescaled expansions by proposing   Pad\'e approximants for $(1-\eta)^dZ(\eta)$. A rescaled Pad\'e approximant constructed from the first $n+m+1$ virial coefficients will be denoted here as $Z_{[n,m]}^{\text{MSAV}}(\eta)$. Thus,
\beqa
Z_{[1,1]}^{\text{MSAV}}(\eta)&=&\frac{1}{(1-\eta)^d}\nn
&&\times
\frac{b_2-d+[d(d+1)/2+b_2(b_2-d)-b_3]\eta}{b_2-d-[b_3-b_2 d+d(d-1)/2]\eta}.\nn
\label{x7}
\eeqa
By construction, $Z_{[n,0]}^{\text{MSAV}}(\eta)=Z_{[n,0]}^{\text{BC}}(\eta)$.
\subsubsection{{ Song--Mason--Stratt approximation}}
Using simple arguments, Song et al.\cite{SMS89,SM90} proposed the following
generalization to $d$ dimensions of the celebrated Carnahan--Starling (CS) EOS
for three-dimensional hard spheres:\cite{CS69}
\beq
Z_{\text{SMS}}(\eta)=1+b_2 \eta\frac{1+(b_3/b_2-d)\eta}{(1-\eta)^d}.
\label{x6}
\eeq
\subsubsection{{ Luban--Michels approximation}}
On a different vein, Luban and Michels\cite{LM90} wrote the
compressibility factor as
\beq
Z_{\text{LM}}(\eta)=1+b_2\eta\frac{1+\left[b_3/b_2-
\zeta(\eta)b_4/b_3\right]\eta}{1-\zeta(\eta)(b_4/b_3)\eta+
\left[\zeta(\eta)-1\right]
(b_4/b_2)\eta^2}.
\label{a4}
\eeq
The knowledge of the function $\zeta(\eta)$ is  equivalent to that of
$Z(\eta)$.
However, $\zeta(\eta)$ focuses on the high density behavior of the EOS, since
Eq.\ (\ref{a4}) is consistent with the exact first four
virial coefficients, regardless of the choice of  $\zeta(\eta)$. The
approximation $\zeta(\eta)=1$ is equivalent to assuming a  Pad\'e
approximant  $Z_{[2,1]}(\eta)$. Instead, Luban and Michels observed that
the computer simulation data for $d=2$--5 favor a {\em linear\/} approximation
$\zeta(\eta)=a+b\eta$, with coefficients obtained by a least-square fit to the
simulation results for each dimensionality.

\subsubsection{{ Percus--Yevick theory}}

 It is noteworthy  that the Percus--Yevick (PY) integral equation can be
 solved analytically in odd dimensions, as first pointed out by Freasier and
 Isbister\cite{FI81} and, independently, Leutheusser.\cite{L84}
The
latter concluded that, in general, the problem reduces to an algebraic equation
of degree $d-3$. Following his procedure, however, we find that for $d \geq 9$
this is not so (see the Appendix) and our
calculations suggest that such
degree should rather be $2^{(d-3)/2}$ for $d \ge 3$. In any case, in five
dimensions one has to
 deal with a quadratic equation\cite{FI81,L84} and
 explicit expressions for the virial and compressibility routes to the EOS can
 be obtained.\cite{BMC99}
A simple analysis of the solution for $d=5$, that as far as we know has not
been carried out before, shows that the
 virial route incorrectly gives a negative value for $B_6$:
 $B_6^{\text{PY-v}}/B_2^5=-\frac{2999}{16^5}\simeq -0.00286$. The
 compressibility route yields  $B_6^{\text{PY-c}}/B_2^5=\frac{12233}{8
 \times16^5}\simeq 0.00146$, while the correct value is
 $B_6/B_2^5\simeq 0.00094$.\cite{CM03b,CM03c} Both routes consistently predict
 that $B_8$ is negative, with subsequent coefficients alternating in sign.  On
 the other hand, the virial route gives values for the magnitude of $B_n$
 ($n\geq 8$) increasingly larger than  the compressibility route:
 $B_n^{\text{PY-v}}/B_n^{\text{PY-c}}\approx 0.66+0.77n$. The alternating
 character of the virial series predicted by the PY equation for $d=5$ is due
 to a branch singularity located on the \textit{negative} real axis at
 $\eta_{\text{branch}}=-(9-5\sqrt{3})/6\simeq -0.0566243$. The radius of
 convergence $\eta_{\text{conv}}^{\text{PY}}\simeq 0.0566243$ of the PY
 solution for $d=5$ is larger than the  value $2^{-5}=0.03125$ extrapolated
 from the  radius $\lim_{d\to\infty}\eta_{\text{conv}}=2^{-d}$, but is close to
 the estimate $\eta_{\text{conv}}\simeq 0.052$ made by Clisby and McCoy on the
 basis of Monte Carlo evaluation of sets of Ree--Hoover diagrams.\cite{CM03b,
 CM03c} All these estimates are sensibly smaller than the packing fraction
 $\eta_f=0.19$ at which freezing occurs for $d=5$.\cite{MT84,LM90,FSL02}

\subsubsection{{ Generalized Carnahan--Starling approximation}}

As is well known, the CS EOS for three-dimensional hard spheres can be
interpreted as a weighted average between the PY virial and compressibility
routes:
\beq
Z_{\text{CS}}(\eta)=\alpha Z_{\text{PY-c}}(\eta)+(1-\alpha)
Z_{\text{PY-v}}(\eta),
\label{a3}
\eeq
where $\alpha=\frac{2}{3}$. Given that the PY equation can be solved for odd
dimensions, it is then natural to speculate about whether or not  the
prescription (\ref{a3}), with an adequate choice of the mixing parameter
$\alpha$, keeps being reliable for $d>3$, even though the internal
inconsistency between both routes seems to increase dramatically with the
dimensionality.\cite{FI81} In the five-dimensional case, one of us\cite{S00}
showed that the choice $\alpha=\frac{3}{5}$ leads to values of $Z_{\text{CS}}$ in
excellent agreement with computer simulations.\cite{MT84} This suggested that
the choice $\alpha=(d+1)/2d$ might provide a good description for $d\geq 3$.
Note that, while Eqs.\ (\ref{x6}) and (\ref{a3}) coincide at $d=3$, they
differ for $d>3$, so they generalize the original CS EOS along different
directions.
An alternative generalization of the CS EOS was made by Gonzalez et al.\
\cite{GGS91} They proposed a simple ansatz for the direct correlation
function $c(r)$, which reduced exactly to the PY theory for $d=1$ and $d=3$
and gave results very close to the PY theory for other dimensions. Their
generalized CS EOS consisted of a weighted average between the virial and
compressibility routes obtained from their theory with a mixing parameter
$\alpha=2(2d-1)/5d$.

\subsection{{ Aim of the paper}}

The aim of this paper is {threefold. First, we present the
explicit  solution to the PY equation in the case of a
seven-dimensional hard-sphere fluid following the procedure introduced by
Leutheusser.\cite{L84} This allows us to derive the EOS of the fluid both
through the virial and the compressibility routes, as well as to analyze the
behavior of the virial coefficients stemming out of them.
{ As we will see, the singularity closest to the origin is again a branch point on the negative real axis, so the radius of convergence of the PY virial series is $\eta_{\text{conv}}\simeq 0.0100625$. We conjecture that this value might be close to the (unknown) exact radius. Moreover, a Carnahan--Starling-like equation of state of the form (\ref{a3}) with $\alpha=\frac{5}{6}$ is proposed.}
Secondly, we
provide molecular dynamics results for the compressibility factor. { To the best of our knowledge, this is the first time that simulation results are presented for hard hyperspheres in seven dimensions. The twenty  densities considered
range from the dilute regime ($\rho\sigma^7=0.1$ or $\eta=0.0037$) to our estimated  freezing point ($\rho\sigma^7\simeq 1.95$ or $\eta\simeq 0.072$). Finally,} we perform a comparison between different proposals for the EOS of a
seven-dimensional hard-sphere fluid with the simulation data. { We observe that the proposals (\ref{x5}) and (\ref{x6}) (which do not have any empirical parameter), (\ref{a4}) (which contains two fitting parameters), and (\ref{a3}) (with one fitting parameter) reproduce fairly well the simulation data.}}

The paper is organized as follows. {In Section \ref{sec2} we provide the
 solution of the PY equation for a seven-dimensional hard-sphere fluid
as well as the analysis of the virial coefficients arising from the derivation
of the EOS using the virial and the compressibility routes. This is followed
in Section \ref{sec3} by a description of the molecular dynamics simulation
that was carried out to obtain the compressibility factor of the fluid. The
results of the simulation are then used to assess the merits of various
proposals that have been made in the literature for the EOS. The
paper is closed in Section \ref{sec4} with further discussion of the results
and some concluding remarks.}

\section{Solution of the Percus--Yevick equation for a seven-dimensional
hard-sphere fluid\label{sec2}}
As mentioned in Sec.\ \ref{sec1}, the solution to the PY equation for  hard
hyperspheres with $d=\text{odd}$ reduces to an algebraic equation of degree
$2^{(d-3)/2}$.
{ The case $d=5$, which yields a quadratic equation, has been analyzed by several authors.\cite{FI81,L84,GGS90,BMC99,S00}}
The highest dimensionality for which the algebraic
problem certainly lends itself to an analytic solution is $d=7$.
{ A sketch of the general solution and some details of the particular cases $d=7$ and $d=9$ are provided in the Appendix.
It is shown there that the solution of the PY equation for seven-dimensional hard hyperspheres is given by the physical solution to the quartic equation (\ref{b14}).}
In the Appendix it is also shown that for $d=9$
the resulting algebraic equation is of eighth degree.

{ A study of the solutions of Eq.\ (\ref{b14}) shows that in the interval $0.446469\lesssim\eta<1$ the four roots are real.} On the other
hand, for $0\leq \eta\lesssim 0.446469$  two of the roots become complex
conjugates and only { the other} two roots remain real, the physical one being finite in
the limit $\eta\to 0$. { The explicit solution to Eq.\ (\ref{b14})} involves the term $\left[P_4(\eta)P_6(\eta)\right]^{1/2}$, where
$P_4(\eta)=1 + 94\eta  + 202{\eta }^2 + \frac{1360}{3}{\eta }^3 + 50{\eta }^4$
and $P_6(\eta)=1 + 99\eta  - \frac{307}{8}{\eta }^2 - \frac{339}{4}{\eta }^3 -
\frac{2762}{3}{\eta }^4 +
\frac{695}{2}{\eta }^5 + \frac{5575}{108}{\eta }^6$. As a consequence, { the solution}
 possesses branch points at the
zeroes of $P_4(\eta)$ and $P_6(\eta)$. The  zero of $P_4(\eta)$ closest to the
origin is
$\eta_{\text{branch}}'\simeq -0.0108868$, while that of $P_6(\eta)$ is
$\eta_{\text{branch}}\simeq -0.0100625$.
{ Therefore, the
radius of convergence of the virial series for a seven-dimensional hard-sphere
fluid described by the PY approximation is $\eta_{\text{conv}}^{\text{PY}}=|\eta_{\text{branch}}|\simeq 0.0100625$.}

 { Table \ref{table2} gives the first few values of the PY virial coefficients obtained from the virial and the compressibility routes.}
{ As far as we know, the exact values $B_n^{\text{ex}}$ of the virial coefficients of seven-dimensional hard spheres are known up to $n=6$ only.\cite{CM03b,CM03c}
They are  listed in Table \ref{table2} as well, which also gives}
the CS-like values $B_n^{\text{CS}}/B_2^{n-1}$, where
$B_n^{\text{CS}}=\alpha B_n^{\text{PY-c}}+(1-\alpha) B_n^{\text{PY-v}}$ with
the simple choice $\alpha=\frac{5}{6}$. Note that the choice $\alpha\simeq 0.6$
would make $B_4^{\text{CS}}\simeq B_4^{\text{ex}}$, whereas the choice
$\alpha\simeq 0.7$ would make $B_5^{\text{CS}}\simeq B_5^{\text{ex}}$ and
$B_6^{\text{CS}}\simeq B_6^{\text{ex}}$. However, comparison with molecular
dynamics simulations (see Section \ref{sec3}) favors $\alpha\simeq 0.8$.
\begin{table*}[t]
\caption{\label{table2}
Values of $B_{n}/B_2^{n-1}$ for $n=3$--8, according to the virial route of the PY approximation, the compressibility route of the PY approximation, the CS-like approximation (\protect\ref{a3}) with $\alpha=\frac{5}{6}$, and the known exact results. \protect\cite{CM03b,CM03c}}
\begin{ruledtabular}
\begin{tabular}{ccccc}
$n$&$B_n^{\text{PY-v}}/B_2^{n-1}$&$B_n^{\text{PY-c}}/B_2^{n-1}$ & $B_n^{\text{CS}}/B_2^{n-1}$ &$B_n^{\text{ex}}/B_2^{n-1}$ \\
\hline
3& $0.2822265625$&$0.2822265625$ &$0.2822265625$&$0.2822265625$\\
4& $-7.499694824\times 10^{-3}$  &$2.155081431\times 10^{-2}$&$1.670906279\times 10^{-2}$&$9.873(4)\times 10^{-3}$\\
5& $1.235022893\times{10}^{-2}$  &$5.116807918\times 10^{-3}$&$6.322378086\times 10^{-3}$&$7.071(7)\times 10^{-3}$\\
6& $-8.177005666\times{10}^{-3}$ &$-1.865328120\times 10^{-3}$&$-2.917274378\times 10^{-3}$&$-3.52(2)\times 10^{-3}$\\
7& $6.553131160\times{10}^{-3}$  &$1.384246670\times 10^{-3}$&$2.245727418\times 10^{-3}$&\\
8& $-5.762797816\times{10}^{-3}$ &$-1.078783146\times 10^{-3}$&$-1.859452258\times 10^{-3}$&\\
\end{tabular}
\end{ruledtabular}
\end{table*}

\begin{figure}[tbp]
\includegraphics[width=.90 \columnwidth]{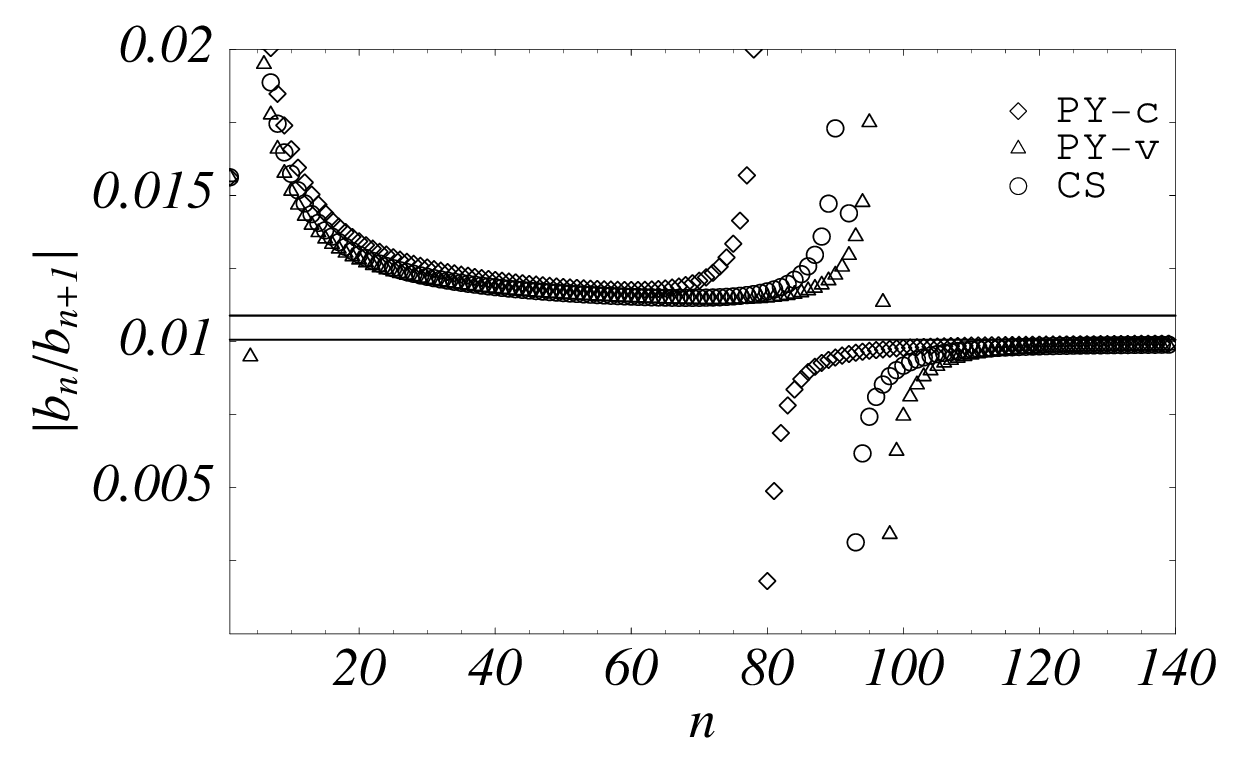}
\caption{\label{ratio} Plot of the ratios
$|b_n^{\text{PY-c}}/b_{n+1}^{\text{PY-c}}|$ (diamonds),
$|b_n^{\text{PY-v}}/b_{n+1}^{\text{PY-v}}|$ (triangles), and
$|b_n^{\text{CS}}/b_{n+1}^{\text{CS}}|$ (circles). The horizontal lines
correspond to the apparent radius of convergence
$\eta_{\text{conv}}'\simeq 0.0108868$ and to the true radius  of convergence
$\eta_{\text{conv}}^{\text{PY}}\simeq 0.0100625$.}
\end{figure}

{}From Table \ref{table2} we observe that the virial route of the PY
approximation incorrectly yields a negative value for the fourth virial
coefficient (which actually becomes negative for $d\geq 8$,
\cite{CM03a,CM03b,CM03c}) while the compressibility route predicts the
correct sign.\cite{note3}
We have computed $B_n^{\text{PY-v}}$ and $B_n^{\text{PY-c}}$ for values of
$n$ much larger than those displayed in Table \ref{table2}. The results
indicate that $\text{sgn}\left(B_n^{\text{PY-v}}\right)=(-1)^{n+1}$ for
$5\leq n\leq 97$ but $\text{sgn}\left(B_n^{\text{PY-v}}\right)=(-1)^{n}$ for
$n\geq 98$; analogously, $\text{sgn}\left(B_n^{\text{PY-c}}\right)=(-1)^{n+1}$
for $5\leq n\leq 80$ but $\text{sgn}\left(B_n^{\text{PY-c}}\right)=(-1)^{n}$
for $n\geq 81$. Therefore,
both routes synchronize their signs for $5\leq n\leq 80$ and again for
$n\geq 98$. This peculiar behavior of the alternating character of the virial
series  seems to be  a consequence of the proximity between the two branch
point singularities closest to the origin,
$\eta_{\text{branch}}\simeq -0.0100625$ and
$\eta_{\text{branch}}'\simeq -0.0108868$, both located on the negative real
axis. To confirm this interpretation, we plot in Fig.\ \ref{ratio} the ratios
$|b_n^{\text{PY-v}}/b_{n+1}^{\text{PY-v}}|$,
$|b_n^{\text{PY-c}}/b_{n+1}^{\text{PY-c}}|$, and
$|b_n^{\text{CS}}/b_{n+1}^{\text{CS}}|$.
Recall that the radius of convergence of the virial series is
$\eta_{\text{conv}}=\lim_{n\to\infty}|b_n/b_{n+1}|$.
Figure \ref{ratio} shows that for $n\lesssim 50$ the ratio $|b_n/b_{n+1}|$
seems to converge from above to the \textit{apparent} radius of convergence
$\eta_{\text{conv}}'=-\eta_{\text{branch}}'\simeq 0.0108868$. However, the
true radius
$\eta_{\text{conv}}^{\text{PY}}=-\eta_{\text{branch}}\simeq 0.0100625$ is
reached from below for $n\gtrsim 100$.

\begin{figure}[tbp]
\includegraphics[width=.90 \columnwidth]{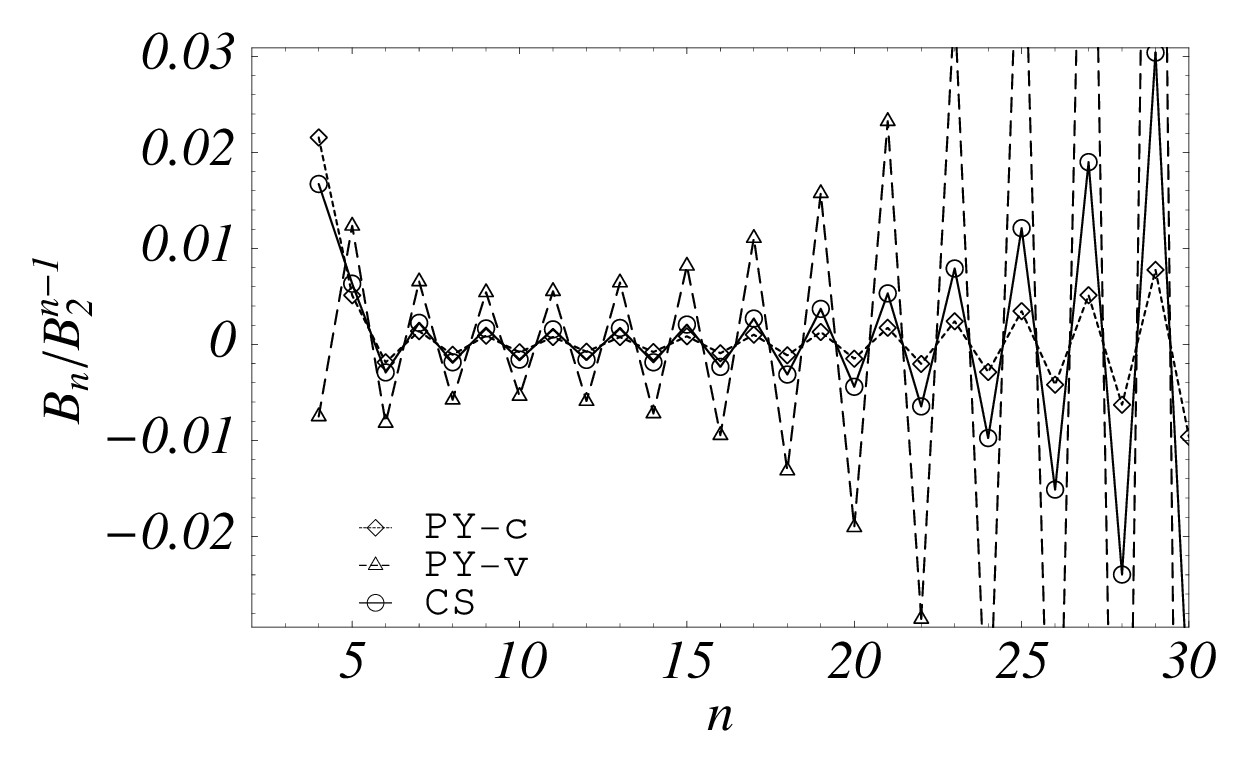}
\caption{\label{virial} Plot of the virial coefficients $B_n^{\text{PY-c}}/B_2^{n-1}$ (diamonds), $B_n^{\text{PY-v}}/B_2^{n-1}$ (triangles), and $B_n^{\text{CS}}/B_2^{n-1}$ (circles).}
\end{figure}

As mentioned in Sec.\ \ref{sec1}, the radius of convergence predicted by the
PY approximation in the five-dimensional case is
$\eta_{\text{conv}}^{\text{PY}}\simeq 0.0566243$. When going to the next odd
dimensionality, the radius of convergence has shrunk to
$\eta_{\text{conv}}^{\text{PY}}\simeq 0.0100625$.
In terms of the scaled density per dimension introduced by Frisch and Percus,
\cite{FP99} the radius of convergence is
$\widehat{\rho}_{\text{conv}}^{\text{PY}}\simeq 1.126$ for $d=5$ and
$\widehat{\rho}_{\text{conv}}^{\text{PY}}\simeq 1.037$ for $d=7$. Therefore,
it nicely tends to converge to the expected value
$\widehat{\rho}_{\text{conv}}=1$ as $d\to\infty$. In addition, the PY value of
$\eta_{\text{conv}}=2^{1-d}B_2\lim_{n\to\infty}|B_{n}/B_{n+1}|$ for $d=7$ is
consistent with estimates obtained from Table XVIII of Ref.\ \onlinecite{CM03b}. By
assuming that the Ree--Hoover ring diagrams dominate for high $d$,\cite{CM03b}
one has
$\eta_{\text{conv}}<2^{-6}B_2|B_{9}/B_{10}|\approx 2^{-6}0.0132/0.0143\simeq 0.014$ for $d=7$, which agrees with the PY value
$\eta_{\text{conv}}^{\text{PY}}\simeq 0.0100625$. Clisby and McCoy's estimate\cite{CM03b}
$\eta_{\text{conv}}\simeq 0.052$ for $d=5$  is also close to the
PY value $\eta_{\text{conv}}^{\text{PY}}\simeq 0.0566243$. All of this leads
us to conjecture that the PY solution gives a fair estimate of the radius of
convergence of the true virial series for high dimensionalities. Pushing this
conjecture even further, we can expect the true radius of convergence to be
due to a singularity (pole or branch point) located on the negative real axis,
so that the virial coefficients alternate in sign beyond a certain order.
Figure \ref{virial} shows the virial coefficients
$B_n^{\text{PY-v}}/B_2^{n-1}$, $B_n^{\text{PY-c}}/B_2^{n-1}$, and
$B_n^{\text{CS}}/B_2^{n-1}$ in the seven-dimensional case. In the spirit of
the above conjecture, one may speculate that the exact values of
$B_n/B_2^{n-1}$ lie in between $B_n^{\text{PY-v}}/B_2^{n-1}$ and
$B_n^{\text{PY-c}}/B_2^{n-1}$, perhaps not far from the interpolated values
$B_n^{\text{CS}}/B_2^{n-1}$. The reduced virial coefficients $B_n/B_2^{n-1}$
start decreasing in magnitude, reach a minimum around $n=10$, and then grow
with $n$.
The fact that the PY solution in the three-dimensional case does not possess
a branch point singularity, so that all the virial coefficients remain
positive, casts some doubts as to whether the true virial series  fails to
converge for densities close to the freezing density $\eta_f\simeq 0.494$.
In any case, the true radius of convergence for $d=3$ cannot be larger than
the crystalline close-packing value
$\eta_{\text{cp}}=\pi\sqrt{2}/6\simeq 0.7405$, while the PY solution has
$\eta_{\text{conv}}^{\text{PY}}=1$.

\section{Molecular dynamics simulations\label{sec3}}

\subsection{Method}

The numerical simulation was implemented by using the same algorithm as
described in Ref.\ \onlinecite{GAH01}, which is also  based on the work of Michels and
Trappeniers\cite{MT84} { and Luban and Michels}\cite{LM90} for four- and five-dimensional hyperspheres.
{ We are not aware of any previous computer simulation of hard hyperspheres of a dimension higher than $d=5$. We have chosen the molecular dynamics method instead of the Monte Carlo method because that gives us the possibility of testing our code by applying it to $d=4$ and $d=5$ and comparing with the results of Refs.\ \onlinecite{MT84,LM90}.}

For our simulations, in order to keep the computing time within reasonable
limits and at the same time being able to examine a wide density range,
the initial configuration is chosen to be the one obtained
by placing $N=64$ hyperspheres in a unitary cell of a $d$-type lattice.
The simulation cell is a hypercube of side $L$ and volume $V=L^d=N/\rho$ and
the minimum image convention and periodic boundary conditions in all directions
have been applied, in the same way as in the 3D case.\cite{AT87}

During the simulations only binary collisions are taken into account, while
collisions between three or four particles are ignored. The collision time for
every pair of particles is calculated and the the smallest value is obtained.
All the particles are moved during this time at constant velocity. The pair of
particles that suffers a collision is treated according to  impulsive
dynamics and the velocities are changed; in this step the hard collisional
virial is calculated. This allows one to evaluate the excess compressibility
factor as
\begin{equation}
Z-1=-\frac{1}{N \langle v^2\rangle \Delta t} \sum_{ij} \mathbf{r}_{ij} \cdot
\Delta \mathbf{v}_i,
\label{c1}
\end{equation}
where  $\langle v^2\rangle$ is the  mean square velocity, $\Delta t$ is the
simulated time, $\mathbf{r}_{ij}$ is the relative position vector between
colliding  particles $i$ and $j$, and  $\Delta \mathbf{v}_i$ the change in
velocity of the particle $i$ on collision.

The equation of state is achieved by changing the diameter $\sigma$ of the
particles, in such a way that the reduced density $\rho^* =\rho \sigma^d$
changes, and letting the system to relax up to an equilibrium pressure.
The errors associated with our calculation were computed following standard
methods for errors in equilibrium averages.\cite{AT87}

{ As mentioned above, before running the program for $d=7$, it  has been  previously validated for $d=4$ and $d=5$,
reproducing the excess compressibility factor obtained by Michels,
Trappeniers, and Luban.\cite{MT84,LM90}}

\subsection{Results}

We have computed the compressibility factor for densities
$0.1\leq \rho^*\leq 1.90$ with a step $\Delta\rho^*=0.1$, as well
as for $\rho^*=1.95$. The simulation data obtained by our molecular
dynamics simulations are listed in Table \ref{table3}. At the largest density
$\rho^*=1.95$ ($\eta=0.0720$) the compressibility factor presents a dramatic drop. We
interpret this as an indication of the freezing transition. Consequently, the
density at which the seven-dimensional fluid of hard spheres freezes can be
estimated as $\rho_f^*\lesssim 1.95$ or, equivalently,
$\eta_f\lesssim 0.072$. {}From Fig.\ 5 of Ref.\ \onlinecite{FSL02} one can observe
that $\ln \eta_f(d)$ is almost a linear function of the dimensionality $d$,
with a slight negative curvature. According to this, knowing the freezing
densities $\eta_f(d)$ and $\eta_f(d+2)$, one can estimate the freezing density
$\eta_f(d+4)$ as $\eta_f(d+4)\lesssim \eta_f^2(d+2)/\eta_f(d)$. Given that
$\eta_f(3)\simeq 0.494$ and $\eta_f(5)\simeq 0.19$, one has $\eta_f(7)\lesssim 0.19^2/0.494\simeq 0.073$, in close agreement with our
estimate.
An independent estimate based on a conjecture by Colot and Baus\cite{CB86}
confirms again this value. These authors suggested that the ratio of length
scales $[\eta_f(d)/\eta_{\text{cp}}(d)]^{1/d}$  is practically independent of
$d$, so that
$\eta_f(d+2)\simeq \eta_{\text{cp}}(d+2)[\eta_f(d)/\eta_{\text{cp}}(d)]^{(d+2)/d}$. The general expression for the
close-packing fraction $\eta_{\text{cp}}(d)$ is not known, but for $d<25$ the values are not far from Blichfeldt's upper estimate\cite{FSL02} $\eta_{\text{cp}}(d)\leq 2^{-d/2}(d+2)/2$.  Using $\eta_f(5)\simeq 0.19$, we get $\eta_f(7)\simeq 0.076$, which is again consistent with our estimate.
\begin{table*}[h]
\caption{\label{table3}
Compressibility factor as a function of $\eta$ { from the simulation data and for different approximations. The numbers in parentheses indicate the statistical error in the last significant digit.}}
\begin{ruledtabular}
\begin{tabular}{cccccccccccc}
$\eta$&$Z_{\text{simul}}$&$Z_{\text{CS}}$&$Z_{\text{PY-v}}$&$Z_{\text{PY-c}}$&$Z_{[4,0]}$&$Z_{[2,2]}$&$Z_{[3,2]}$&$Z_{[2,0]}^{\text{BC}}$&$Z_{[2,2]}^{\text{MSAV}}$
&$Z_{\text{SMS}}$&$Z_{\text{LM}}$ \\
\hline
$0.0037$ & $1.25366(2)$ & $1.25223$ & $1.25192$ & $1.25229$ & $1.25214$ & $1.25214$ & $1.25214$ & $1.25233$ & $1.25214$ & $1.25233$ & $1.25217$ \\
$0.0074$ & $1.5337(1)$ & $1.53751$ & $1.53516$ & $1.53797$ & $1.53687$ & $1.53687$ & $1.53681$ & $1.53829$ & $1.53685$ & $1.53827$ & $1.53725$ \\
$0.0111$ & $1.8646(3)$ & $1.85767$ & $1.84997$ & $1.85921$ & $1.85577$ & $1.85576$ & $1.85534$ & $1.86008$ & $1.85562$ & $1.86003$ & $1.85745$ \\
$0.0148$ & $2.2103(3)$ & $2.21482$ & $2.19691$ & $2.21840$ & $2.21093$ & $2.21094$ & $2.20930$ & $2.22006$ & $2.21033$ & $2.21993$ & $2.21565$ \\
$0.0185$ & $2.6174(2)$ & $2.61124$ & $2.57674$ & $2.61814$ & $2.60499$ & $2.60513$ & $2.60047$ & $2.62071$ & $2.60328$ & $2.62047$ & $2.61524$ \\
$0.0221$ & $3.0650(4)$ & $3.04946$ & $2.99039$ & $3.06128$ & $3.04111$ & $3.04170$ & $3.03080$ & $3.06467$ & $3.03716$ & $3.06423$ & $3.0600$ \\
$0.0258$ & $3.5449(5)$ & $3.53219$ & $3.43890$ & $3.55085$ & $3.52298$ & $3.52481$ & $3.50243$ & $3.55469$ & $3.51502$ & $3.55399$ & $3.55394$ \\
$0.0295$ & $4.0989(7)$ & $4.06234$ & $3.92342$ & $4.09012$ & $4.05480$ & $4.05948$ & $4.01769$ & $4.09372$ & $4.04037$ & $4.09264$ & $4.10109$ \\
$0.0332$ & $4.7013(5)$ & $4.64302$ & $4.44519$ & $4.68258$ & $4.64133$ & $4.65183$ & $4.57911$ & $4.68483$ & $4.61713$ & $4.68326$ & $4.70519$ \\
$0.0369$ & $5.389(1)$ & $5.27757$ & $5.00555$ & $5.33198$ & $5.28785$ & $5.30924$ & $5.18944$ & $5.33130$ & $5.24971$ & $5.32910$ & $5.36929$ \\
$0.0406$ & $6.051(1)$ & $5.96955$ & $5.60592$ & $6.04228$ & $6.00015$ & $6.04073$ & $5.85164$ & $6.03659$ & $5.94307$ & $6.03358$ & $6.09511$ \\
$0.0443$ & $6.8179(6)$ & $6.72276$ & $6.24777$ & $6.81775$ & $6.78456$ & $6.85731$ & $6.56896$ & $6.80433$ & $6.70273$ & $6.80033$ & $6.88228$ \\
$0.0480$ & $7.6325(1)$ & $7.54121$ & $6.93269$ & $7.66292$ & $7.64794$ & $7.77255$ & $7.34490$ & $7.63837$ & $7.53489$ & $7.63317$ & $7.72727$ \\
$0.0517$ & $8.5133(2)$ & $8.42922$ & $7.66233$ & $8.58259$ & $8.59769$ & $8.80333$ & $8.18328$ & $8.54278$ & $8.44645$ & $8.53613$ & $8.62220$ \\
$0.0554$ & $9.4294(3)$ & $9.39134$ & $8.43841$ & $9.58192$ & $9.64172$ & $9.97083$ & $9.08829$ & $9.52186$ & $9.44519$ & $9.51348$ & $9.55355$ \\
$0.0591$ & $10.492(1)$ & $10.4324$ & $9.26275$ & $10.6664$ & $10.7885$ & $11.3020$ & $10.0645$ & $10.5801$ & $10.5398$ & $10.5697$ & $10.5009$ \\
$0.0628$ & $11.570(3)$ & $11.5577$ & $10.1372$ & $11.8417$ & $12.0469$ & $12.8317$ & $11.1168$ & $11.7224$ & $11.7400$ & $11.7096$ & $11.4365$ \\
$0.0664$ & $12.694(1)$ & $12.7725$ & $11.0638$ & $13.1142$ & $13.4266$ & $14.6056$ & $12.2507$ & $12.9537$ & $13.0569$ & $12.9381$ & $12.3252$ \\
$0.0701$ & $13.907(3)$ & $14.0828$ & $12.0446$ & $14.4904$ & $14.9374$ & $16.6847$ & $13.4722$ & $14.2794$ & $14.5029$ & $14.2607$ & $13.1265$ \\
$0.0720$ & $9.03944(6)$ & $14.7756$ & $12.5559$ & $15.2196$ & $15.7454$ & $17.8637$ & $14.1178$ & $14.9794$ & $15.2786$ & $14.9589$ & $13.4810$ \\
\end{tabular}
\end{ruledtabular}
\end{table*}

 Table \ref{table3} also gives some theoretical values: the PY predictions
 $Z_{\text{PY-v}}$ and $Z_{\text{PY-c}}$, the CS-like interpolation (\ref{a3})
 with $\alpha=\frac{5}{6}$, the truncated virial expansion $Z_{[4,0]}$, the
 Pad\'e approximants $Z_{[2,2]}$ and  $Z_{[3,2]}$ [the three latter being
 obvious extensions of the approximations (\ref{x2})--(\ref{x4})], the rescaled
 virial expansion $Z_{[2,0]}^{\text{BC}}$ defined by Eq.\ (\ref{x5}),  the
 rescaled Pad\'e approximant $Z_{[2,2]}^{\text{MSAV}}$ defined by a natural extension
 of Eq.\ (\ref{x7}), the SMS approximation (\ref{x6}), and the LM proposal
 (\ref{a4}).

Although the knowledge of the sixth virial coefficient $B_6$ would allow one
to consider the truncated series $Z_{[5,0]}$, it is not included in Table
\ref{table3} because it turns out to be clearly inferior to $Z_{[4,0]}$. This is a
consequence of the fact that $B_6<0$, so that $Z_{[5,0]}<Z_{[4,0]}$, while
for small and moderate  densities $Z_{[4,0]}<Z_{\text{simul}}$. This is a strong
indication that the unknown seventh virial coefficient $B_7$ must be positive.
Among the different Pad\'e approximants that can be constructed from the
knowledge of the first six virial coefficients, the best agreement with the
simulation data is presented by $Z_{[2,2]}$ for $\rho^*\lesssim 1.4$ ($\eta\lesssim 0.0517$) and
by $Z_{[3,2]}$ for $\rho^*\gtrsim 1.4$ ($\eta\gtrsim 0.0517$) . It is interesting to note that
both Pad\'e approximants have poles on the negative real axis (at
$\eta\simeq -0.079$ in the case of  $Z_{[2,2]}$ and at $\eta\simeq -0.025$ in
the case of  $Z_{[3,2]}$), so that the extrapolated virial coefficients have
alternating signs.
Paradoxically, while the  rescaled expansion $Z_{[2,0]}^{\text{BC}}$
incorporates the first three virial coefficients only, it exhibits a better
agreement with simulation than those rescaled expansions that can be constructed with the first
four, five, or six virial coefficients, so the latter are not included in
Table \ref{table3}.
Analogously, the best performance among the rescaled Pad\'e approximants
corresponds to $Z_{[2,2]}^{\text{MSAV}}$.
Interestingly, the SMS proposal [cf. Eq.\ (\ref{x6})] and the approximation
$Z_{[2,0]}^{\text{BC}}$  yield practically equivalent results. The difference
between both EOS is
\beqa
Z_{[2,0]}^{\text{BC}}(\eta)-Z_{\text{SMS}}(\eta)&=&\frac{\eta^3}{(1-\eta)^7}\left(35-35\eta\right.\nn
&&\left.+21\eta^2-7\eta^3+\eta^4\right).
\label{8}
\eeqa
This corresponds to a relative difference smaller than 0.14\% for the density
range considered in the simulations.

\begin{figure}[tbp]
\includegraphics[width=.90 \columnwidth]{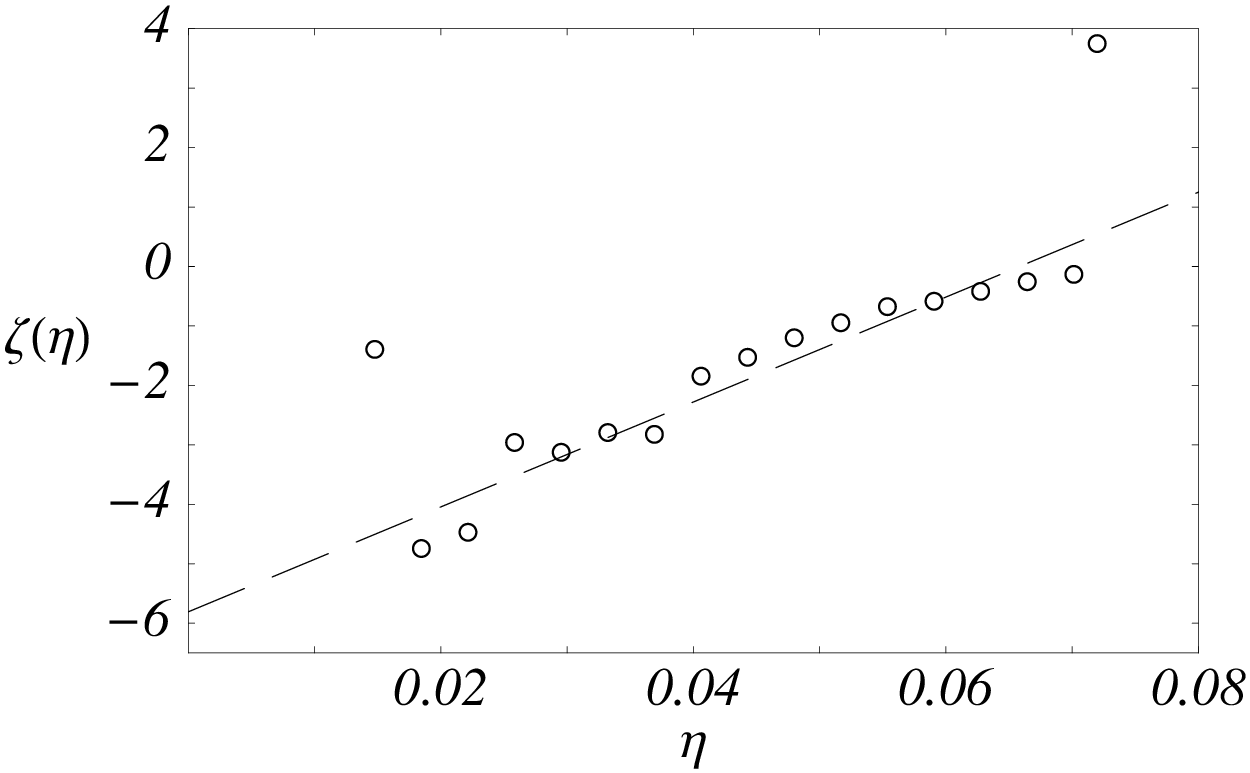}
\caption{\label{zeta} Plot of the simulation values of the function
$\zeta(\eta)$ defined by Eq.\ (\protect\ref{a4}). The dashed line is the
linear fit (\protect\ref{c2}).}
\end{figure}

Two of the theoretical EOS included in Table \ref{table3}, namely $Z_{\text{CS}}$ and $Z_{\text{LM}}$, have an empirical
character. The proposal
(\ref{a3}) is based on the observation that the two PY routes tend to bracket
the simulation data, as happens in the three-dimensional\cite{CS69} and
five-dimensional\cite{S00} cases.
We have found that the value  $\alpha=\frac{5}{6}$ of the parameter is the
simplest rational number that makes $Z_{\text{CS}}$ reproduce fairly well the
simulation values.
In the case of the Luban--Michels EOS (\ref{a4}) one fits $\zeta(\eta)$ to a
linear function. Figure \ref{zeta} shows the simulation values of $\zeta(\eta)$.
As in the five-dimensional case,\cite{LM90} $\zeta(\eta)$ is an increasing
function of $\eta$, while it is a decreasing function for $d=2$--4. A linear
fit in the interval $0.5\leq \rho^*\leq 1.9$ ($0.0185\leq\eta\leq 0.0701$) yields
\beq
\zeta(\eta)=-5.81+88.2\eta.
\label{c2}
\eeq
The column labeled $Z_{\text{LM}}$ in Table \ref{table3} has been evaluated
using the fit (\ref{c2}).
On the other hand, our simulation data { in Fig.\ \ref{zeta}} seem to indicate a negative curvature
of $\zeta(\eta)$.

\begin{figure}[tbp]
\includegraphics[width=.90 \columnwidth]{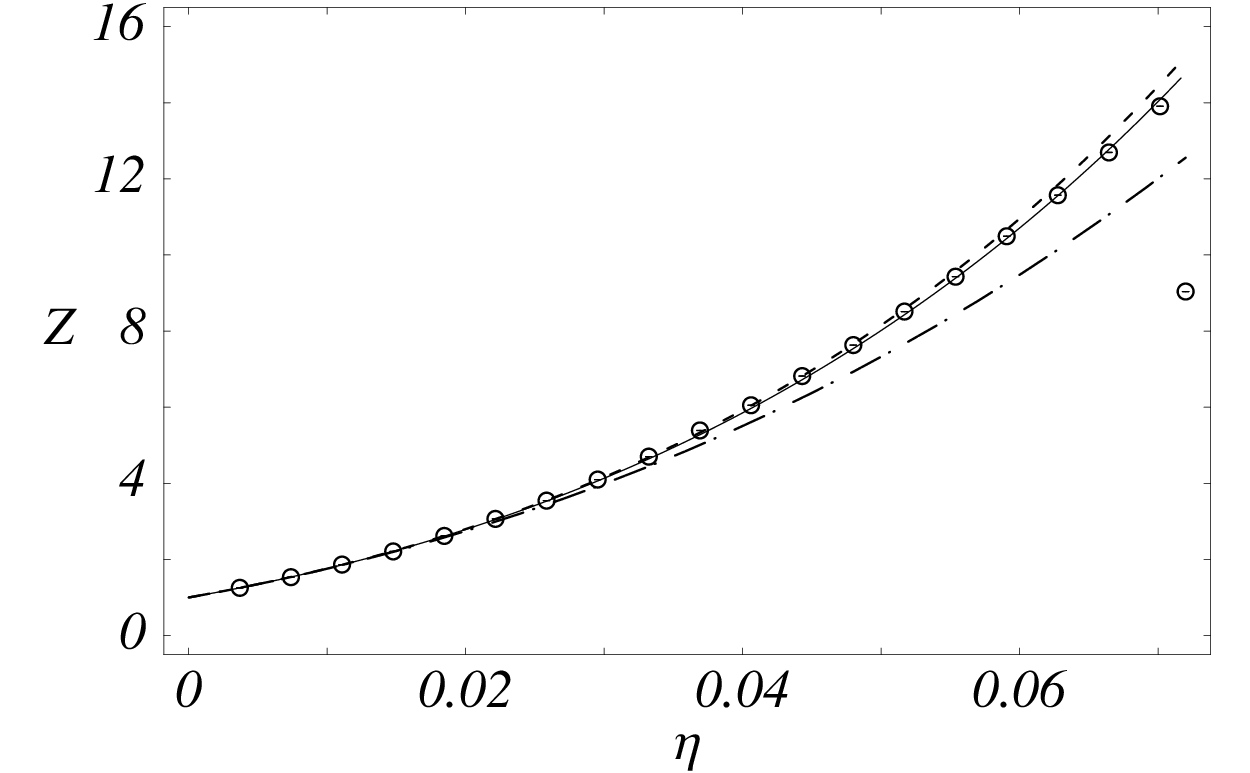}
\caption{\label{EOS} Compressibility factor as a function of the packing
fraction. The circles represent simulation data, the dashed line
represents $Z_{\text{PY-c}}$, the dash-dotted line represents
$Z_{\text{PY-v}}$, and the solid line represents the CS-like interpolation
(\protect\ref{a3}) with $\alpha=\frac{5}{6}$.}
\end{figure}

Table \ref{table3} shows that up to $\rho^*=0.8$ ($\eta=0.0295$) all the different
theoretical results tabulated, including the simple truncated virial expansion
$Z_{[4,0]}$, behave relatively well. For larger densities, $Z_{[4,0]}$  tends
to overestimate the simulation data, while the Pad\'e approximants $Z_{[2,2]}$
and  $Z_{[3,2]}$ tend to  underestimate them. The best global agreement is
presented by $Z_{\text{CS}}$,  $Z_{\text{LM}}$,
$Z_{[2,0]}^{\text{BC}}$, and $Z_{\text{SMS}}$. This is especially noteworthy
in the case of the two latter approximations, since they do not contain
fitting parameters and, moreover, only the knowledge of the first three virial
coefficients is exploited. This contrasts with $Z_{\text{LM}}$, which includes
the fourth virial coefficient and contains two fitting parameters.
On the other hand, $Z_{\text{CS}}$ belongs in a different class of
approximations. Given the involved algebraic structure of the PY solution,
$Z_{\text{CS}}$ does not intend to represent a practical recipe to the EOS of
a seven-dimensional hard-sphere fluid. Instead, its role is to highlight the
fact that the two PY routes keep bracketing the simulation data, so that an
interpolation between them with a density-independent parameter $\alpha$ is
rather accurate, as graphically illustrated in Fig.\ \ref{EOS}. This gives
some confidence on the expectation that some of the analytical properties of
the PY solution (e.g., alternating character of the virial series, branch
points located on the negative real axis, \ldots) may shed light on the true
behavior of the exact series.

\section{Discussion\label{sec4}}

The results of the previous sections deserve further discussion. To begin with,
to our knowledge this is the first time that a molecular dynamics simulation has
been carried out on a seven-dimensional hard-sphere fluid. The simulation strategy
that we adopted implied a compromise between computer process time and density range
to be explored and the outcome is rather encouraging. The availability of simulation
data for the EOS of the fluid allowed us to locate the freezing transition and also
to assess the merits and limitations of various proposals that have been made in the
literature for the compressibility factor of hard hyperspheres. From this analysis it
is clear that even simple approximations such as  $Z_{[2,0]}^{\text{BC}}$ and $Z_{\text{SMS}}$ do
a reasonably good job and that, as it occurs in other dimensionalities, the virial and
compressibility routes to the EOS in the PY approximation keep bracketing the simulation
data, so that a Carnahan--Starling-like recipe of the form of Eq.\ (\ref{a3}) turns out to be
rather accurate. However the parameter $\alpha$ seems not to follow a simple relation as
the ones suggested by  Gonz\'alez et al.\cite{GGS91,GGS92} or Santos.\cite{S00}

We also presented the explicit solution of the PY equation for a hard-sphere fluid
in 7D.
 Such a solution
allowed us to carry out an analysis of the virial coefficients arising both in the
virial and in the compressibility routes and to determine the radius of convergence
of both virial series. The results indicate some peculiar behavior of the virial
coefficients with the virial route incorrectly predicting a negative fourth virial
coefficient. The radius of convergence of the virial series is due to a singularity
(branch point) located on the negative real axis and therefore what one has is an alternating
series. Because of the good agreement between our value of the radius of convergence
of the virial series and other independent estimates and the similar results obtained
for $d=5$, it is tempting to conjecture that the PY solution for even higher
dimensionalities should provide a rather accurate estimate of the radius of convergence
of the true virial series and that it is the existence of singularities on the negative
real axis (either poles or branch points) which determines such radius.

As a final point it is worth commenting that in this case our analysis was facilitated
by the fact that we could combine both the analytical and the simulation results. And
due to the common features such as the freezing transition that hard-core systems in
different dimensionalities share, the expectation and the hope is that the present results
shed some more light on the thermodynamic properties of such systems.
As far as the high dimensionality limit is concerned, our results provide some support to the scenario of Frisch and Percus mentioned in the Introduction in the following sense. The solution to the PY equation predicts an alternating virial series. Further, the values of the scaled density $\widehat{\rho}$ that one obtains for the radius of convergence ($\widehat{\rho}=1.13$ for $d=5$, $\widehat{\rho}=1.04$ for $d=7$, and the number $\widehat{\rho} \simeq 1.02$ coming out of our preliminary
calculations for $d=9$) are consistent with a limiting value of $\widehat{\rho}=1$ for $d\to\infty$. Also, the fluid range in  $d=7$ is reasonably well accounted for by the first three or four virial coefficients so that it is conceivable that for infinite dimensionality only the second virial coefficient will be the dominant term.

\acknowledgments
M.R. acknowledges the financial support of CONACyT through project
No.\ I38644-E.
M.L.H. acknowledges the hospitality of Universidad Complutense de Madrid (Spain),
where the final version of the paper was prepared, as well as the financial
support of DGAPA-UNAM during his sabbatical stay in Madrid.
The research of
A.S. has been partially supported by the Ministerio de
Ciencia y Tecnolog\'{\i}a
 (Spain) through grant No.\ BFM2001-0718 and  by the European
Community's Human Potential Programme under contract HPRN-CT-2002-00307,
DYGLAGEMEM.

\appendix
\section{ Solution of the Percus--Yevick equation for hard hyperspheres}
For simplicity, in the remainder of this { Appendix}
we set $\sigma=1$.

In the Percus--Yevick approximation, the structure factor $S(q)$ of a
hard-sphere fluid in $d=2k+1$ dimensions is
\begin{equation}
S(q)=\frac{1}{\widetilde{Q}(q) \widetilde{Q}(-q)},
\label{bb1}
\end{equation}
where
\begin{equation}
\label{bb2}
\begin{array}{l}
\widetilde{Q}(q)=1-\lambda \int_{0}^{1} dr\, e^{iqr} Q(r),\\
 \lambda\equiv (2\pi)^{k}\rho=2^{2k}(2k+1)!!\,\eta ,
\end{array}
\end{equation}
$Q(r)$ having the form
\beq
Q(r)=
\left\{
\begin{array}{ll}
\sum_{m=0}^{k} Q_m (r-1)^{m+k}, & 0\leq r\leq 1,\\
0,& r\geq 1.
\end{array}
\right.
\label{b1}
\end{equation}
The $k+1$ coefficients $\{Q_m\}$ are functions of the density  determined by
the two linear equations
\beq
(-1)^k=-k! 2^k Q_k + \lambda \sum_{m=0}^{k}(-1)^m \frac{Q_m}{k+m+1}, \quad k\geq 0,
\label{b2}
\eeq
\beqa
(-1)^k&=&-(k-1)! 2^{k-1} Q_{k-1}\nn
&& + \lambda \sum_{m=0}^{k} (-1)^m \frac{Q_m}{k+m+2},\quad k\geq 1,
\label{b4}
\eeqa
plus the $k-1$ nonlinear equations
\beqa
Q^{(2m+1)}(0)&=&\frac{1}{2} \lambda (-1)^{m+1}\left[Q^{(m)}(0)\right]^2\nn
&& -
\lambda \sum_{\nu=0}^
{m-1}(-1)^{\nu}Q^{(\nu)}(0)Q^{(2m-\nu)}(0), \nn
&& 0\leq m\leq k-2.
\label{b5}
\eeqa
Here $Q^{(\nu)}(r)$ represents the $\nu$-th derivative of the
function $Q(r)$. For $k=0$ ($d=1$), Eq.\ (\ref{b2}) gives the exact solution
for hard rods. For $k=1$ ($d=3$), Eqs.\ (\ref{b2}) and (\ref{b4}) are
sufficient to find the solution of the PY equation. However, for $k\geq 2$
($d\geq 5$) one needs in addition Eq.\ (\ref{b5}), so that the problem reduces
to solving an algebraic equation which, as we will argue below,
is likely to be of degree $2^{k-1}=2^{{(d-3)/2}}$.

In the limit $\eta\to 0$, it is easy to verify that
\beq
\begin{array}{l}
\lim_{\eta\to 0}Q_m=(-1)^{k+1}\frac{2^{-m}}{k!}\binom{k}{m}, \\ \lim_{\eta\to 0}Q(r)=(-1)^{k+1}\frac{2^{-k}}{k!}(r^2-1)^k,
\end{array}
\label{b15}
\eeq
\beq
\begin{array}{l}
\lim_{\eta\to 0}Q^{(2m)}(0)=(-1)^{m+1}2^{-k}\frac{(2m)!}{m!(k-m)!},\\ \lim_{\eta\to 0}Q^{(2m+1)}(0)=0.
\end{array}
\label{b16}
\eeq
In general, one can expand the coefficients $Q_m$ in powers of $\eta$:
\beq
Q_m(\eta)=\sum_{n=0}^\infty Q_{m,n}\eta^n,
\label{b17}
\eeq
where $Q_{m,0}$ is given by the first equation of (\ref{b15}). Of course, the
full nonlinear dependence of the coefficients $Q_m(\eta)$ can be obtained from
the solution to the set of equations (\ref{b2})--(\ref{b5}), either
analytically ($k\leq 3$) or numerically ($k\geq 4$).

Once one has determined the function $Q(r)$, the structural properties of the
fluid are given by Eqs.\ (\ref{bb1}) and (\ref{bb2}). In particular, the long
wavelength limit of the structure factor { and the contact value of the radial distribution function are, respectively,}
\begin{equation}
S(q=0)=\frac{1}{\left[k!2^kQ_k\right]  ^2} ,
\label{b6}
\end{equation}
\beq
g(1^+)=(-1)^{k+1}k! Q_0.
\label{b7}
\eeq
The virial route to the EOS is given by
\beq
Z=1+2^{d-1}\eta g(1^+),
\label{b8}
\eeq
while the compressibility route is
\begin{equation}
\chi\equiv k_B T \left(\frac{\partial\rho}{\partial p}\right)_T=S(q=0).
\label{b9}
\end{equation}
Inserting the expansion (\ref{b17}) into Eqs.\ (\ref{b8}) and (\ref{b9}) we
get the virial coefficients along both routes:
\beq
\begin{array}{l}
b_{n+2}^{\text{PY-v}}=2^{2k}(-1)^{k+1}k! Q_{0,n},\\ b_{n+1}^{\text{PY-c}}=2^{2k}(k!)^2\frac{1}{n+1}\sum_{m=0}^nQ_{k,m} Q_{k,n-m}.
\end{array}
\label{b18}
\eeq

\subsection{The case $d=7$}
{ Now we particularize to the seven-dimensional case ($k=3$),  the unknowns being } $Q_m$, $m=0,1,2,3$.
Since the two nonlinear equations (\ref{b5}) involve the derivatives
$Q^{(m)}\equiv Q^{(m)}(0)$, it is more advantageous to work with the set
$\{Q^{(m)}\}$ rather than with the set $\{Q_m\}$. The latter can be expressed
in terms of the former as
\beq
\left(
\begin{array}{c}
Q_0\\
Q_1\\
Q_2\\
Q_3
\end{array}
\right)=-
\left(
\begin{array}{cccc}
 20&10&2 &\frac{1}{6}\\
45&25&\frac{11}{2}& \frac{1}{2}\\
36&21&5&\frac{1}{2}\\
10&6&\frac{3}{2}&\frac{1}{6}
\end{array}
\right)\cdot
\left(
\begin{array}{c}
Q^{(0)}\\
Q^{(1)}\\
Q^{(2)}\\
Q^{(3)}
\end{array}
\right).
\label{b10}
\eeq
Equations (\ref{b2}) and (\ref{b4}), plus Eq.\ (\ref{b5}) with $m=0$ yield
\begin{widetext}
\beq
Q^{(1)}=-3360\eta {Q^{(0)}}^2,
\label{b11}
\eeq
\beq
Q^{(2)}=-
\frac{ 1 + 96Q^{(0)}\left\{ 1 -
         5\eta \left[3 + 112Q^{(0)}\left( 3 - 10\eta  \right)  \right]  \right\}   }
    {8\left( 1- \eta  \right)},
\label{b12}
\eeq
\beq
Q^{(3)}=
\frac{8 - 15 \eta  + 192 Q^{(0)} \left\{ 2 -
       \eta  \left[ 53 + 280 Q^{(0)} {\left( 3 - 10 \eta  \right) }^2 - 100 \eta  \right]  \right\} }
    {8 {\left( 1- \eta  \right) }^2}
\label{b13}
\eeq
{ Thus, the parameters $Q^{(1)}$, $Q^{(2)}$, and $Q^{(3)}$ are given as explicit quadratic functions of $Q^{(0)}=Q(0)$. Finally, insertion} of Eqs.\ (\ref{b11})--(\ref{b13}) into Eq.\ (\ref{b5}) with $m=1$ leads to the quartic equation
\beqa
&&8 - 15\eta  + 192 Q^{(0)} \left\{ 2 -
     \eta  \left\{ 88 - 135 \eta  +1960  Q^{(0)}
         \left[ 3 - 4 \eta  \left[ 9 - 10 \eta  +
              240  Q^{(0)} \left( 1 - \eta  \right)
\right.\right.\right.\right.
\nn
&&\left.\left.\left.\left.
 \times\left( 3 - 10 \eta\left( 1 + 84  Q^{(0)} \left( 1 - \eta  \right)  \right)    \right)
              \right]  \right]  \right\}  \right\}=0.
\label{b14}
\eeqa
\end{widetext}
Although an explicit expression exists for the physical root of Eq.\
(\ref{b14}), it is of course too cumbersome and will be omitted here. \cite{newnote2}

Table \ref{table1} shows the first few coefficients $Q_{m,n}$. The exact
values are rational numbers, but they become more and more involved as the
order $n$ increases and so they are expressed in real form in Table
\ref{table1}. {}From Eq.\ (\ref{b18}) we can obtain the virial coefficients
corresponding to the virial and the compressibility routes. The first few
values  are listed in Table \ref{table2}.
\begin{table*}[t]
\caption{\label{table1}
{ Values of the coefficients $Q_{m,n}$, defined by Eq.\ (\protect\ref{b17}), for $n=0$--6.}}
\begin{ruledtabular}
\begin{tabular}{ccccc}
$n$&$Q_{0,n}$&$ Q_{1,n}$ & $Q_{2,n}$ &$ Q_{3,n}$\\
\hline
0&0.166666667&0.250000000&0.125000000&0.020833333\\
1&3.010416667&7.888020833&5.812500000&1.333333333\\
2& $-5.119791667$ &$-59.313802083$&$-41.072916667$&$-6.541666667$\\
3& $5.395897352\times 10^2$ &$4.247567790\times 10^3$&$3.201932292\times 10^3$&$6.540590278\times 10^2$\\
4& $-2.286456505\times 10^4$ &$-2.488562475\times 10^5$&$-1.884527380\times 10^5$&$-3.841568446\times 10^4$\\
5& $1.172728501\times{10}^6$ &$1.635350292\times{10}^7$&$1.241794603\times{10}^7$&$2.538798289\times{10}^6$\\
6& $-6.600274174\times{10}^7$ &$-1.143524052\times{10}^9$&$-8.688923174\times{10}^8$&$-1.778764760\times{10}^8$\\
\end{tabular}
\end{ruledtabular}
\end{table*}

\subsection{The case $d=9$}
We will now sketch the result for the case $d=9$ following the same procedure.
For $k=4$, the set $\{Q_m\}$ can be expressed in terms of the set
$\{Q^{(m)}\}$ as
\beq
\left(
\begin{array}{c}
Q_0\\
Q_1\\
Q_2\\
Q_3\\
Q_4
\end{array}
\right)=
\left(
\begin{array}{ccccc}
 70&35&\frac{15}{2}&\frac{5}{6}&\frac{1}{24}\\
224&119&27&\frac{19}{6}& \frac{1}{6}\\
280&154&\frac{73}{2}&\frac{9}{2}& \frac{1}{4}\\
160&90&22&\frac{17}{6}&\frac{1}{6}\\
35&20&5&\frac{2}{3}&\frac{1}{24}
\end{array}
\right)\cdot
\left(
\begin{array}{c}
Q^{(0)}\\
Q^{(1)}\\
Q^{(2)}\\
Q^{(3)}\\
Q^{(4)}
\end{array}
\right).
\label{N1}
\eeq
In addition, the fifth derivative $Q^{(5)}$ is a linear combination of $\{Q_m\}$ and hence of the first four derivatives:
\beq
Q^{(5)}=-20\left(336 Q^{(0)}+210 Q^{(1)}+60 Q^{(2)}+10 Q^{(3)}+Q^{(4)}\right).
\label{N2}
\eeq
The nonlinear equations (\ref{b5}) with $m=0,1,2$ allow one to express the odd derivatives in terms of the even ones as
\beq
Q^{(1)}=-\frac{\lambda}{2}{Q^{(0)}}^2,
\label{N3}
\eeq
\beq
Q^{(3)}=\frac{\lambda^3}{8}{Q^{(0)}}^4-\lambda {Q^{(0)}}{Q^{(2)}},
\label{N4}
\eeq
\beq
Q^{(5)}=-\frac{\lambda^5}{16}{Q^{(0)}}^6+\frac{\lambda^3}{2}{Q^{(0)}}^3{Q^{(2)}}-\frac{\lambda}{2}{Q^{(2)}}^2 -\lambda {Q^{(0)}}{Q^{(4)}},
\label{N5}
\eeq
where $\lambda=241\,920\eta$. Next, insertion of Eqs.\ (\ref{N3}) and (\ref{N4}) into the linear equations (\ref{b2}) and (\ref{b4}) yields ${Q^{(2)}}$ and ${Q^{(4)}}$ as \textit{nonlinear} functions of ${Q^{(0)}}$.
Finally, by equating the right-hand sides of Eqs.\ (\ref{N2}) and (\ref{N5}) one gets a closed algebraic equation of eighth degree for ${Q^{(0)}}$. A preliminary analysis of this equation indicates that its physical solution possesses a branch point at $\eta_{\text{branch}}\simeq -0.0023945$, so that the radius of convergence of the PY virial series would be
$\eta_{\text{conv}}=|\eta_{\text{branch}}|\simeq 0.0023945$.

We have checked that for $d=11$ the resulting equation is of degree 16. Therefore, it seems plausible that in the general case $d=2k+1$ the degree of the equation for ${Q^{(0)}}$ is $2^{k-1}=2^{(d-3)/2}$.


\begin{thebibliography}{29}
\bibitem{newnote2}
A code written in Mathematica which provides the explicit expression for this physical root is available from any of the authors upon request.
\end{thebibliography}

\end{document}